\documentclass{article}
\usepackage[english]{babel}
\usepackage[T2A]{fontenc}
\usepackage[cp1251]{inputenc}
\usepackage[dvips]{graphicx}
\usepackage{amsmath}
\usepackage{amssymb}
\usepackage{latexsym}
\begin{document}
\begin{flushright}
Journal-Ref: Astronomy Letters, 2013, Vol. 39, No. 1, pp. 26-37 
\end{flushright}
\begin{center}
\LARGE {\bf Asymmetric Illumination of a Circumbinary Disk in the Presence
of a Low-Mass Companion}\\

\vspace{1cm}
\renewcommand{\thefootnote}{$\ast$}
\small{\copyright 2012 г.} \large  {\bf  T.\,Demidova$^{1}$\footnote{e-mail: proxima1@list.ru}, V.\,Grinin$^{1,2}$,
N.\,Sotnikova$^{1,2}$}
\end{center}

\normalsize
\begin{center}
1 - Pulkovo Astronomical Observatory, Russian Academy of Sciences,\\ Pulkovskoe shosse 65, St. Petersburg,
                                         196140 Russia, \\

2 - Sobolev Astronomical Institute, St. Petersburg State University,\\ Universitetskii pr. 28, St. Petersburg, 198504 Russia,
\end{center}
\normalsize
\begin{center}
Received May 18, 2012
\end{center}
\begin{abstract}
The model of an young star with a protoplanetary disk and a low-mass
companion ($q \leq 0.1$), which is moving on a circular orbit, inclined to the disk plane, is
considered. The hydrodynamic models of such a system were calculated by SPH method. 
It was shown the distortions in the disk, caused by the orbital
motion of the companion, lead to the strong dependence of illumination
conditions of the disk on the azimuth (because of extinction between the star and the disk 
surface) and, therefore, it leads to the appearance of large-scale asymmetry in images of disks. The
calculations showed the dependence of the illumination on the azimuth was stronger in the
central part of the disk than on the periphery. Bright and dark domains are located not symmetry with respect 
to the line of nodes. The sizes and locations of the domains are depended on the model parameters as well as on 
the phase of the orbital period. The bright and dark domains do not follow the
companion, but they make small amplitude oscillations with respect to some direction.
The properties of the model, which were written above, open new opportunities of searching low-mass
companions in the vicinity of young stars. The stars with protoplanetary disks, which are observed face-on
or under the small inclination angle $i$, are the best ones for this purpose.

\end{abstract}

\clearpage
\twocolumn
\renewcommand{\thefootnote}{}

\section*{INTRODUCTION}

\renewcommand{\thefootnote}{\arabic{footnote}}
Many young stars are known to be surrounded
by circumstellar gas-dust disks radiating in the
infrared (IR) and submillimeter wavelength ranges
(see, e.g., the review by Natta et al. (2000)). In the
overwhelming majority of cases, observations provide
only information about the integrated (i.e., from
the entire disk) IR radiation (Mendoza 1966; Cohen 1975). In the visual spectral range,
the circumstellar disks radiate very little, mainly
through the scattered radiation of their central stars\footnote[1]{UX Ori stars at eclipses (Grinin et al. 1991) and young
objects with edge-on circumstellar disks constitute an exception.}.Nevertheless, most of the circumstellar disk images
known to date have been obtained precisely in the
optical spectral range with the Hubble Space Telescope
by the method of coronagraphy (McCaughrean
and O'Dell 1996; McCaughrean et al. 2000; Grady
et al. 2000; Krist et al. 2000; Clampin et al. 2003; and
others). Among them, there are disks seen edge-on or at a small angle to the line of sight (Burrows
et al. 1995; McCaughrean and O'Dell 1996;
Stapelfeldt et al. 1998; and others) and disks seen
nearly face-on (Grady et al. 2000; Krist et al. 2000;
and others). Many of the disks have an asymmetric
shape that is often explained by anisotropic scattering
of stellar radiation by circumstellar dust grains. This
mechanism is efficient in the cases where the disk is
inclined to the plane of the sky.

However, the nature of the asymmetry in circumstellar
disk images is not exhausted by this mechanism
alone. The asymmetry of the disk surface
brightness can also be caused by an anisotropic illumination
of the disk by the central star, for example,
due to the presence of spots on the star. This model
was considered by Wood and Whitney (1998) in connection
with the detection of variability in the image of
HH30's protoplanetary disk by Burrows et al. (1996).
It is quite real, because the rotational modulation of
the brightness of young stars due to a nonuniform
surface brightness is a well-known observational fact
(see, e.g., Vrba et al. 1986).

Another cause of the anisotropic disk illumination
can be the absorption of stellar radiation by circumstellar matter if it is distributed nonuniformly in azimuth.
Hydrodynamic calculations show (Artymowicz
and Lubow 1996; Larwood and Papaloizou 1997;
Sotnikova and Grining 2007; Hanawa et al. 2010;
Kaigorodov et al. 2010) that such conditions can arise
in young binary systems accreting matter from the
remnants of a protostellar cloud (below referred to as a
CB (circumbinary) disk). Under the action of periodic
perturbations induced by the orbital motion of the
companion, spiral density waves and gas flows propagating
into the system's inner regions are formed
in the CB disk. If such a system is observed at a
small angle to the disk plane, then, as our calculations
showed (Sotnikova and Grinin 2007; Demidova
et al. 2010a, 2010b; Grinin et al. 2010), the motion of
large-scale structures in it can cause great circumstellar
extinction variations and, as a consequence, a
large-amplitude brightness modulation of the system.
It turned out than a noticeable (in amplitude) photometric
effect could be caused by the orbital motion of
a low-mass companion or a protoplanet (Demidova
et al. 2010b) and that this effect is enhanced if the
companion's orbit is inclined to the disk plane (Grinin
et al. 2010).

It should be noted that the situation where the
orbit of the companion (planet) does not coincide with
the midplane of the common disk or the equatorial
plane of the central star is not something exceptional.
This can result from a nonaxisymmetric distribution
of specific angular momentum in the protostellar
cloud from which the star and protoplanetary disk are
born. A star whose rotation axis may not coincide
with the rotation axis of the protoplanetary disk is
formed during the gravitational contraction of such a
cloud. This possibility has been discussed in recent
years (see, e.g., Bate et al. 2010) in connection with
the unexpected results of spectroscopic observations
of stars with exoplanets during the transit of a planet
over the stellar disk. They showed that the orbital
planes of several exoplanets do not coincide with the
equatorial plane of the stars (H\'ebrard et al. 2009;
Winn et al. 2009a, 2009b; Pont et al. 2010; Narita
et al. 2011; Brown et al. 2012). Previously, Burrows
et al. (1995) found that the inner part of the circumstellar
disk around $\beta$ Pic Pic is inclined by several degrees
to its outer part and hypothesized that this inclination
was caused by the motion of a planet whose orbit
is also inclined to the disk plane. The corresponding
model was developed by Mouillet et al. (1997)
and Larwood and Papaloizou (1997), and the planet
itself has also been discovered recently (Lagrange
et al. 2009; Chauvin et al. 2012). The same picture
is observed for the Herbig Ae star CQ Tau: the inner
part of its circumstellar disk is inclined approximately
by $30^\circ$ to the disk periphery (Eisner et al. 2004;
Chapillon et al. 2008).

In this paper, we investigate the influence of hydrodynamic
perturbations in such systems on the
extinction variations between the central star and disk
surface. The goal of our calculations is to determine
the disk illumination conditions at various distances
from the center and its dependence on azimuth.\\
\clearpage

\section*{THE METHOD OF CALCULATIONS}

We consider the model of a binary system that
consists of a central star and a low-mass companion ($q = m_2:m_1 \le 0.1$), moving in a circular orbit
of radius $a$. The orbital plane is assumed to be
inclined to the CB disk midplane. The mass of the
star is taken to be $2\,M_\odot$ this value roughly corresponds
to the mass of a typical Herbig Ae star); the orbital
period of the companion is 1 year.

As in our previous papers, the 3D hydrodynamic
calculations were performed by the SPH (Smoothed
Particle Hydrodynamics) method in the isothermal
approximation. The implementation of this method
is described in detail in Sotnikova (1996). One of the
main hydrodynamic parameters, the effective viscosity
of the disk, was determined via the dimensionless
speed of sound $c$, expressed in fractions of the Keplerian
velocity in the companion's orbit. This parameter
was varied within the range $c = 0.001-0.08$. The parameters of the problem also include the mass
accretion rate onto the binary components
$\dot{M_a}$ and the opacity $\kappa$, per gram of circumstellar matter (as in
our previous papers, $\kappa$ was taken to be $250$ cm$^2$/g (Natta and Whitney 2000), which corresponds to the
optical properties of the circumstellar dust in young
circumstellar disks in the visual spectral range). The
dust-to-gas mass ratio was taken to be the same as
that, on average, in the interstellar medium, $1:100$.

The calculations were performed using from $60000$ to $2\times10^5$
test particles in a region of radius $10\,a$. Our
calculations showed that after 30 rotations the system
finishes the relaxation stage (with the formation of
a central gap) and the distribution of SPH particles
reflects best the behavior of the matter in the CB disk.
The disk model calculated in this way was smoothed
over the cells of a 3D mesh (with a step of $0.1\,a$). This procedure allowed the influence of random
fluctuations in the distribution of SPH particles to
be reduced while retaining all details of the flow
structure.

As an example, Fig. 1 shows the distribution of
SPH particles in one of the calculated models (before
the smoothing procedure described above). We
see that the deviations from azimuthal symmetry are
greatest near the inner boundary of the CB disk (the
disk is puffed up). At distances of the order of five orbital
radii, the distribution of matter in the CB disk is
already azimuthally symmetric, and such a disk differs
little in structure from the disk around a single star of
the same mass. To investigate the influence of density
waves and nonuniform structure of the inner region of
the CB disk on the illumination of its outer layers, the
model distribution of particles was smoothly extended
to distances of $50\,a$ by using the standard model of a
flared disk:
\begin{equation}
h(r)= h_0\Big(\frac{r}{r_0}\Big)^\beta,
\end{equation}
where $2\,h(r)$ --- is the geometrical thickness of the disk
at distance $r$ from the symmetry axis, the parameter $\beta = 5/4$ (Kenyon and Hartmann 1987), and the initial
values of $h_0 = a$, and the radius $r_0 = 5a$.\\

\subsection*{\emph{Calculation of the Disk Illumination}}

We calculated the disk surface illumination from
the formula
\begin{equation}
 F(r,\phi) = \frac{L_{\ast}e^{-\tau(r,\phi)}}{4\pi(r^2+h^2)}\sin\gamma,
\end{equation}
where $L_{\ast}$ --- is the luminosity of the central star, $\tau(r,\phi)$
--- is the optical depth of the dust layer between the star
and an arbitrary point on the disk at distance $r$ from
the symmetry axis and azimuth $\phi$, $\gamma$ --- is the angle
between the vector connecting the star and the point
on the disk surface and the tangent to the disk surface
at this point (for details, see Tambovtseva et al. 2006).

When calculating $\tau(r,\phi)$ we used the same
method as in our previous papers (see, e.g.,Demidova
et al. 2010a): having compared the accretion rate
of test particles onto the binary components with
the accretion rate specified as a parameter of the
problem (in the range $\dot{ M_a} = 10^{-9} -10^{-11} M_\odot yr^{-1}$), we determine the test particle mass $m_d$. Then, $\tau(r,\phi) = m_d \cdot n(r,\phi) \cdot k/S$, where $n(r,\phi)$ is the number
density of test particles in the column between the
star and a point on the disk surface with coordinates $r$ и $\phi$, $S$ is the cross section of this column.\\

\subsection*{\emph{Allowance for the Scattered Radiation}}

The direct stellar radiation is the main disk surface
illumination source. This radiation does not penetrate
into the shadow zone because of strong absorption
on the way, in the perturbed disk region, and it is
illuminated only by the radiation scattered by circumstellar
dust. To estimate the contribution from this
radiation, we calculated the intensity of the scattered
radiation for several models in the single-scattering approximation. This approximation is applied in the
cases (see, e.g., Tambovtseva et al. 2006) where the
single-scattering albedo differs noticeably from unity,
because under these conditions the role of multiple
scatterings is minor.

Below, we used the same scheme of calculations
as in the above paper (Tambovtseva et al. 2006):
the inner CB disk region ($r < 5a$) was divided into, cells, each being considered as a scattering element
with density $\rho_i = m_d \cdot n_i$, where $n_i$ s the number of
particles in the $i$th cell. For each surface element of
the CB disk, the scattered light was calculated as the
sum of the fluxes scattered by each cell toward the
chosen disk surface element. We took into account
the absorption of radiation by dust both on the way
from the star to the scattering volume and on the way
from it to the illuminated disk surface element:

\begin{equation}
 F_{sc}(r,\phi) = k_{sc}\sum \limits_{i}\frac{L_{\ast}}{4\pi d_1^2}\frac{\rho_i}{4\pi d_2^2}
 f(\theta_i)e^{-\tau_i}\cos\xi_i.
\end{equation}
Here, $d_1$ is the distance from the primary component
to the scattering cell, $d_2$ is the distance from the
scattering cell to the unit CB disk surface element
illuminated by the scattered radiation, $k_{sc}$ is the scattering
coefficient, $\tau_i = \tau_1+\tau_2$ is the sum of the optical
depths of dust on the way from the primary component
to the scattering cell ($\tau_1$) and from the scattering
cell to the surface element on the outer CB disk ($\tau_2$), $\xi_i$ is the angle between the direction of the scattered
radiation from the center of the scattering cell to the
surface element on the CB disk under consideration
and the normal to the disk surface at this point. In our
calculations, we used the Henyey-Greenstein phase
function:

\begin{equation}
f(\theta) = \frac{1}{4\pi}\frac{1-g^2}{(1+g^2-2g\,\cos{\theta})^{3/2}},
\end{equation}
where $\theta$ is the scattering angle, $g$ is the asymmetry
factor (taken to be $0.5$). The single-scattering albedo
was taken to be $0.4$, which roughly corresponds to the
optical properties of circumstellar dust in the $V$ band (Natta and Whitney 2000). Taking this into account,
the scattering coefficient is $k_{sc} = 100 $cm$^2$/g.\\

\section*{RESULTS}

We calculated the disk surface illumination by the
method described above for several models. The input
model parameters are the orbital inclination to the
disk plane $\alpha$, the component mass ratio $q$, and the
disk viscosity parameter $c$. Figure 2 presents the
results of our calculations for two models differing
by the orbital inclination:  $\alpha = 10^\circ$ and
$\alpha = 30^\circ$. We see
that, in both cases, the disk illumination is highly
nonuniform in azimuth due to the absorption of the
direct stellar radiation in the perturbed region (near
the inner disk boundary). The degree of deformation
of the inner CB disk increases with $\alpha$ causing the
area of the shadow from it on the disk to increase. A
nontrivial peculiarity in the disk brightness distribution
attributable to the distribution of matter in the
perturbed CB disk region is that the \emph{line separating
the bright and dark regions on the disk does not
coincide with the line of nodes}. For quantitative
estimations, we determined the location of this line
in such a way that the ratio of the integrated illuminations
of the disk regions on different sides of this
line was maximal. Our calculations showed that the
inclination of this line to the line of nodes changes
with model parameters within a comparatively narrow
range: $\phi_l \approx 50-70^\circ$. In particular, in the models
with $\alpha = 10^\circ$ the angle $\phi_l = 57^\circ$, with $\alpha = 30^\circ$ --- $\phi_l =
59^\circ$ (Fig. 2).

As can be seen from Fig. 3, the location of the
shadow zone relative to the line of nodes in the models
considered is almost independent of the companion's
orbital inclination. The illumination minimum
is reached in a region with an azimuthal angle $\phi_{sh} \approx 330^\circ$.
Since the minimum illumination is due to the
maximum absorption of radiation in the perturbed
disk region, the result obtained suggests that the
region of maximum disk deformation does not coincide
with the region in which the companion rises to
the maximum height above the CB disk plane ($\phi = 270^\circ$).
Such a distribution of matter in the inner disk
stems from the fact that the secondary component
entrains part of the matter once it has passed the
point of greatest distance from the CB disk plane,
which causes the CB disk thickness in this direction
to increase. Other things being equal, the larger the
viscosity, the higher the efficiency of this process (see
below).

The deformation of the inner CB disk region also
depends on the companion's mass. The greater the
value of $q$, the larger the amplitude of the perturbations
in the disk and the stronger the asymmetry
in its illumination. In this case, the position of the
illumination minimum still corresponds to $\phi_{sh} \approx 330^\circ$ (Fig. 4).

Viscosity affects the illumination conditions in the
following way: the CB disk becomes denser and thinner
with decreasing viscosity. As a result, the area
of the shadow on the disk is reduced. For example,
in the models with a ``cold'' disk ($c \le 0.03$)  the disk
is illuminated uniformly already at $R \ge 20 a$, while in
the models with с $c = 0.05$ (at the same value of $q$) the
shadow zone extends to $R = 30a$. The location of the
line separating the bright and dark regions on the disk
also depends on viscosity and varies within the range
$\phi_l=50-65^\circ$.  In addition, the azimuthal angle of the
illumination minimum in the shadow zone increases
with viscosity (Fig. 5). This suggests that the disk
matter is entrained more actively by the secondary
component during its orbital motion and shields the
star more strongly with increasing viscosity.

Figures 6 and 7 show the two limiting situations
considered in our calculations where a noticeable
asymmetry in the CB disk illumination is still possible:
the model with a small orbital inclination to the
disk plane ($\alpha = 3^\circ$)  and the model in which the mass
of the perturbing body is of the order of several Jupiter
masses ($q = 0.003$). Under these conditions, the
perturbations in the disk induced by the companion's
orbital motion are small. Nevertheless, as can be seen
from Figs. 6 and 7, an asymmetry in the disk illumination
due to an asymmetric distribution of absorbing
matter is clearly seen in the central disk region in
these cases as well. Our calculations showed that
the degree of asymmetry in the disk illumination is
more sensitive to variations in $\alpha$, than to variations
in the companion's mass. Therefore, even a planet
with a mass of several Jupiter masses can produce
a noticeable asymmetry in the protoplanetary disk
illumination if its orbit has an appreciable inclination
to the disk plane ($\alpha \ge 10^\circ$).\\

\subsection*{\emph{Influence of the Scattered Light}}

In addition to the direct radiation, the disk surface
is also illuminated by the scattered light. The
contribution from this radiation on the bright (illuminated)
part of the disk is negligible compared to the
direct stellar radiation. However, the scattered light
in the shadow region makes a tangible contribution
to the disk surface brightness. Where, according to
Figs. 3-5, the illumination by the direction radiation
is close to zero, the disk surface brightness is nonzero
due to the scattered light.

We calculated the disk illumination in the single scattering
approximation for several models. Two of
them are presented in Fig. 8. Analysis of the calculated
models showed that the scattered light reduces
the depth of the shadow zone only slightly in the case
of small disk deformation (at small $\alpha$).
However, at $\alpha=30^\circ$  the contribution from the scattered light
in the shadow zone is more significant. As can be
seen from Figs. 8 and 9, a bright spot emerges in the
shadow zone in the inner CB disk. It appears, because
the inner CB disk is strongly deformed and has
a significant thickness at large angles $\alpha$. Therefore,
the scattered light from the perturbed region on the
CB disk illuminates the outer part of its surface at a
less acute angle than in the models with a lower $\alpha$. As
a result, in the models with strong disk deformation,
the scattered light makes a tangible contribution to
the illumination of the shadow zone.\\

\subsection*{\emph{Orbital Motion of the Companion and the Disk Illumination}}

The changes in disk illumination conditions with
orbital phase are of special interest. Calculations
showed that these changes occur by no means in the
same way as, for example, in the model of a spotted
star, where the regions of maximum and minimum
illumination on the disk move azimuthally, following
the stellar rotation\footnote{The shadow region created by a dusty disk wind from
the secondary component in the models by Tambovtseva
et al. (2006) moves over the disk in the same way.} (Wood and Whitney 1998). In our case, as a result of the companion's orbital
motion, the bright and dark regions execute small
synchronous oscillations on the disk surface, with
their shape changing only slightly. This is clearly seen
from Fig. 10, which shows two disk models. The
companion is at an orbital phase of  $90^\circ$, in one model
and at the opposite end of the orbit, at a phase of $270^\circ$, in the other model. We see that the boundary between
the bright and dark zones in these two limiting situations
changed only slightly. The position angle of the
line separating these two regions changes with orbital
phase, though within a narrow range (Fig. 11).\\

\subsection*{\emph{Asymmetry of Disks During Coronagraphic Observations}}

As has been pointed out in the Introduction, the
method of coronagraphy, whereby the central star
(along with the central region on the disk) is covered,
is commonly used in the observations of circumstellar
disks. We modeled this situation. Figure 12 shows
two disk illumination distributions in the same model.
In one of them, the central region of the system is covered
by an opaque shield. The shield size relative to
the disk size was chosen in such a way that it roughly
corresponded to the conditions of Hubble Space Telescope
observations of the circumstellar disk around
one of the nearest Herbig Ae stars, HD 163296 (see
Wisniewski et al. 2008). We see that, in this case,
the shadow zone is covered almost completely by
the artificial shield. Consequently, the coronagraphic
method of imaging widely used in studying circumstellar
disks can hide valuable information about the
existence of an asymmetry on the disk caused by a
low-mass companion from the observer.

It should be noted that in close binary systems like
those we consider here with an orbital radius of the
order of several AU, the dust in the perturbed region
of the CB disk near its inner boundary makes a major
contribution to the circumstellar extinction. Other
potentially important extinction sources include the
circumstellar disk surrounding the central stars and
the gas flows in the inner region of the system. In the
implementation of the SPH method (with a constant
smoothing length) we used, this region is modeled
rather roughly. Since this region has small sizes in
close binary systems like those considered here, its
contribution to the extinction is negligible. In wider
pairs, the inner region of the system can contain
much dust, and more accurate methods should be
used for its calculations (Hanawa et al. 2010; Fateeva et al. 2011; Sytov et al. 2011; de Val-Borro
et al. 2011).\\

\section*{CONCLUSIONS}

Our calculations show that the perturbations in
the circumstellar disk of a young star induced by the
orbital motion of a low-mass companion whose orbit
is inclined to the disk plane can cause a noticeable
asymmetry in the disk illumination. It differs fundamentally
in properties from what is given by the
previously considered models. For example, in the
models of disks with anisotropic scattering of stellar
radiation by dust grains, a disk asymmetry is possible
only when the disk is inclined to the plane of the sky
(Augereau et al. 1999). During face-on observations,
such a disk will be azimuthally uniform. In contrast,
in our case, an asymmetry in the disk illumination
exists at any disk inclination and is best seen during
face-on observations.

In contrast to the models in which the moving
bright and dark regions on the disk follow the rotation
of a spotted star (Wood and Whitney 1998) or the
motion of a companion that is the source of a disk
wind (Tambovtseva et al. 2006), the orbital motion
of the companion in our model \emph{does not cause any
synchronous motion of the shadow on the disk but
leads only to small oscillations of the bright and
dark regions.}. To a first approximation, the boundary
between them retains its location in projection onto
the plane of the sky. In this case, the line separating
the illuminated region and the shadow zone,
\emph{coincide with the line of nodes}. The position angle
of this line measured from the line of nodes is $40$-$60^\circ$ 
(depending on the model parameters).

The amplitude of the perturbations in the disk
induced by the orbital motion of the companion, along
with the parameters of the bright and dark regions
on the disk, depend on the component mass ratio,
viscosity, and orbital inclination. Other things being
equal, the smaller these parameters, the smaller the
shadow zone on the disk. The minimum mass of
the companion at which the asymmetry in the disk
illumination is still noticeable in our calculations is
about six Jupiter masses (Fig. 6).

Allowance for the influence of scattered light
showed that it reduces the depth of the shadow on
the disk without changing the asymmetry in the disk
illumination. The case where the companion's orbit is
highly inclined to the disk plane (Fig. 8) constitutes
an exception. In this case, the scattered radiation forms a bright spot in the shadow region due to strong
deformation of the inner CB disk. This effect was
obtained in the single-scattering approximation and
deserves a more detailed study by solving the 3D
radiative transfer problem.

The degree of asymmetry in the illuminated region
on the disk is maximal in the central part of the disk
with a radius of $20$-$30$ orbital radii and decreases
with increasing distance from the center. During
coronagraphic observations, this region is covered by
a mask (an artificial moon) covering the star. We
expect a big breakthrough in studying the asymmetry
of circumstellar disk images once the ALMA interferometer
has been put into operation. It will operate
in the submillimeter wavelength range and will allow
the central regions of protoplanetary disks to be seen
without any distortions.\\

This work was supported by the Basic Research
Program of the Presidium of the Russian Academy
of Sciences P21 of the Presidium of the Russian
Academy of Sciences, grant NSh 1625.2012.2 Federal
Targeted Program Science and Science Education
for Innovation in Russia 2012-2013.\\

\newpage
\begin{center}
{\Large\bf REFERENCES}
\end{center}
1. P.\,Artymowicz and S.\,H.\,Lubow, Astrophys. J. \textbf{467},
L77 (1996).\\
2. J.\,C.\,Augereau, A.\,M.\,Lagrange, D.\,Mouillet, et al.,
Astron. Astrophys. \textbf{348}, 557 (1999).\\
3. M.\,R.\,Bate, G.\,Lodato, and J.\,E.\,Pringle, Mon. Not.
R. Astron. Soc., \textbf{401}, 1505 (2010).\\
4. D.\,J.\,A.\,Brown, A.\,Collier Cameron, D.\,R.\,Anderson,
et al.,Mon. Not. R. Astron. Soc. Tmp. \textbf{292} (2012)\\
5. C.\,J.\,Burrows, J.\,E.\,Krist, K.\,R.\,Stapelfeldt, and
WFPC2 Investigation Definition Team, Bull. Am. Astron.
Soc. \textbf{27}, 1329 (1995).\\
6. S.\,J.\,Burrows, K.\,R.\,Stapelfeldt, A.\,M.\,Watson, et al.,
Astrophys. J. \textbf{473}, 437 (1996).\\
7. E.\,Chapillon, S.\,Guilloteau, A.\,Dutrey, et al., Astron.
Astrophys. \textbf{488}, 565 (2008).\\
8. G.\,Chauvin, A.-M.\,Lagrange, H.\,Beust, et al.,\\
arXiv1202.2655C (2012).\\
9. M.\,Clampin, J.\,E.\,Krist, D.\,R.\,Ardila, et al., Astron.
J. \textbf{126}, 385 (2003).\\
10. M.\,Cohen, Mon. Not. R. Astron. Soc. \textbf{173}, 279
(1975).\\
11. T.\,V.\,Demidova, V.\,P.\,Grinin, and N.\,Ya.\,Sotnikova,
Astron. Lett. \textbf{36}, 498 (2010).\\
12. T.\,V.\,Demidova, N.\,Ya.\,Sotnikova, and V.\,P.\,Grinin,
Astron. Lett. \textbf{36}, 422 (2010).\\
13. J.\,A.\,Eisner, B.\,F.\,Lane, L.\,A.\,Hillenbrand, et al.,
Astrophys. J. \textbf{613}, 1049 (2004).\\
14. A.\,M.\,Fateeva, D.\,V.\,Bisikalo, P.\,V.\,Kaygorodov, et al.,
Astrophys. Space Sci. \textbf{335}, 125 (2011).\\
15. S.\,A.\,Grady, D.\,Devine, B.\,Woodgate, et al., Astrophys.
J. \textbf{544}, 895 (2000).\\
16. V.\,P.\,Grinin, N.\,N.\,Kiselev, G.\,P.\,Chernova, et al.,
Astrophys. Space Sci. \textbf{186}, 283 (1991).\\
17. V.\,P.\,Grinin, T.\,V.\,Demidova, and N.\,Ya.\,Sotnikova,
Astron. Lett. \textbf{36}, 808 (2010).\\
18. T.\,Hanawa, Y.\,Ochi, and K.\,Ando, Astrophys. J. \textbf{708},
485 (2010).\\
19. G.\,H\'ebrard, F.\,Bouchy, F.\,Pont, et al., in Transiting
Planets, Ed. by F.\,Pont, D.\,Sasselov, and M.\,J.\,Holman
(Cambridge: Cambridge Univ. Press, 2009),
p. 508.\\
20. P.\,V.\,Kaigorodov, D.\,V.\,Bisikalo, A.\,M.\,Fateeva, et al.,
Astron. Rep. \textbf{54}, 1078 (2010).\\
21. S.\,J.\,Kenyon and L.\,Hartmann, Astrophys. J. \textbf{323},
714 (1987).\\
22. J.\,E.\,Krist, K.\,R.\,Stapelfeldt, F.\,M\'enard, et al., Astrophys.
J. \textbf{538}, 793 (2000).\\
23. M.-A.\,Lagrange, D.\,Gratadour, G.\,Chauvin, et al.,
Astron. Astrophys. \textbf{493}, L21 (2009).\\
24. L.\,D.\,Larwood and J.\,C.\,P.\,Papaloizou, Mon. Not.
R. Astron. Soc. \textbf{285}, 288 (1997).\\
25. M.\,J.\,McCaughrean and C.\,R.\,O'Dell, Astron. J. \textbf{111},
1977 (1996).\\
26. M.\,J.\,McCaughrean, K.\,R.\,Stapelfeldt, and
L.\,M.\,Close, in Protostars and Protoplanets
IV, Ed. by V.\,Mannings, A.\,P.\,Boss, and S.\,S.\,Russell
(Univ. of Arizona Press, Tucson, 2000), p. 485.\\
27. V.\,E.\,Mendoza, Astrophys. J. \textbf{143}, 1010
(1966).\\
28. D.\,Mouillet, J.\,D.\,Larwood, J.\,C.\,B.\,Papaloizou, et al.,
Mon. Not. R. Astron. Soc. 292, \textbf{896} (1997).\\
29. N.\,Narita, T.\,Hirano, B.\,Sato, et al., Publ. Astron.
Soc. Jpn. \textbf{63}, 67 (2011).\\
30. A.\,Natta and B.\,Whitney, Astron. Astrophys. \textbf{364}, 633
(2000).\\
31. A.\,Natta, V.\,P.\,Grinin, and V.\,Mannings, in Protostars
and Planets IV, Ed. by V.\,Mannings,
A.\,P.\,Boss, and S.\,S.\,Russell (Tucson: Univ. of Arizona
Press, 2000), p. 559.\\
32. F.\,Pont, M.\,Endl, and W.\,D.\,Cochran, Mon. Not.
R. Astron. Soc. \textbf{402}, L1 (2010).\\
33. A.\,N.\,Rostopchina, Astron. Rep. \textbf{43}, 113 (1999).\\
34. N.\,Ya.\,Sotnikova, Astrofizika \textbf{39}, 259 (1996).\\
35. N.\,Ya.\,Sotnikova and V.\,P.\,Grinin, Astron. Lett. \textbf{33},
594 (2007).\\
36. K.\,R.\,Stapelfeldt, J.\,E.\,Krist, F.\,M\'enard, et al., Astrophys.
J. \textbf{502}, L65 (1998).\\
37. A.\,Yu.\,Sytov, P.\,V.\,Kaigorodov, A.\,M.\,Fateeva, et al.,
Astron. Rep. \textbf{55}, 793 (2011).\\
38. L.\,V.\,Tambovtseva, V.\,P.\,Grinin, and G.\,Weigelt, Astron.
Astrophys. \textbf{448}, 633 (2006).\\
39. M.\,de Val-Borro, G.\,F.\,Gahm, H.\,C.\,Stempels, et al.,
Mon. Not. R. Astron. Soc. \textbf{413}, 2679 (2011).\\
40. F.\,J.\,Vrba, A.\,E.\,Rydgren, P.\,F.\,Chugainov, et al.,
Astrophys. J. \textbf{306}, 199 (1986).\\
41. J.\,Winn, J.\,A.\,Johnson, D.\,Fabrycky, et al., Astrophys.
J. \textbf{700}, 302 (2009a).\\
42. J.\,Winn, A.\,W.\,Howard, J.\,A.\,Johnson, et al., Astrophys.
J. \textbf{703}, 2091 (2009b).\\
43. J.\,P.\,Wisniewski, M.\,Clampin, C.\,A.\,Grady, et al.,
Astrophys. J. \textbf{682}, 548 (2008).\\
44. K.\,Wood and B.\,Whitney, Astrophys. J. Lett. \textbf{506}, L43
(1998).\\

Translated by V. Astakhov

\onecolumn

\begin{figure}[!h]\begin{center}
  \makebox[0.6\textwidth]{\includegraphics[scale=1]{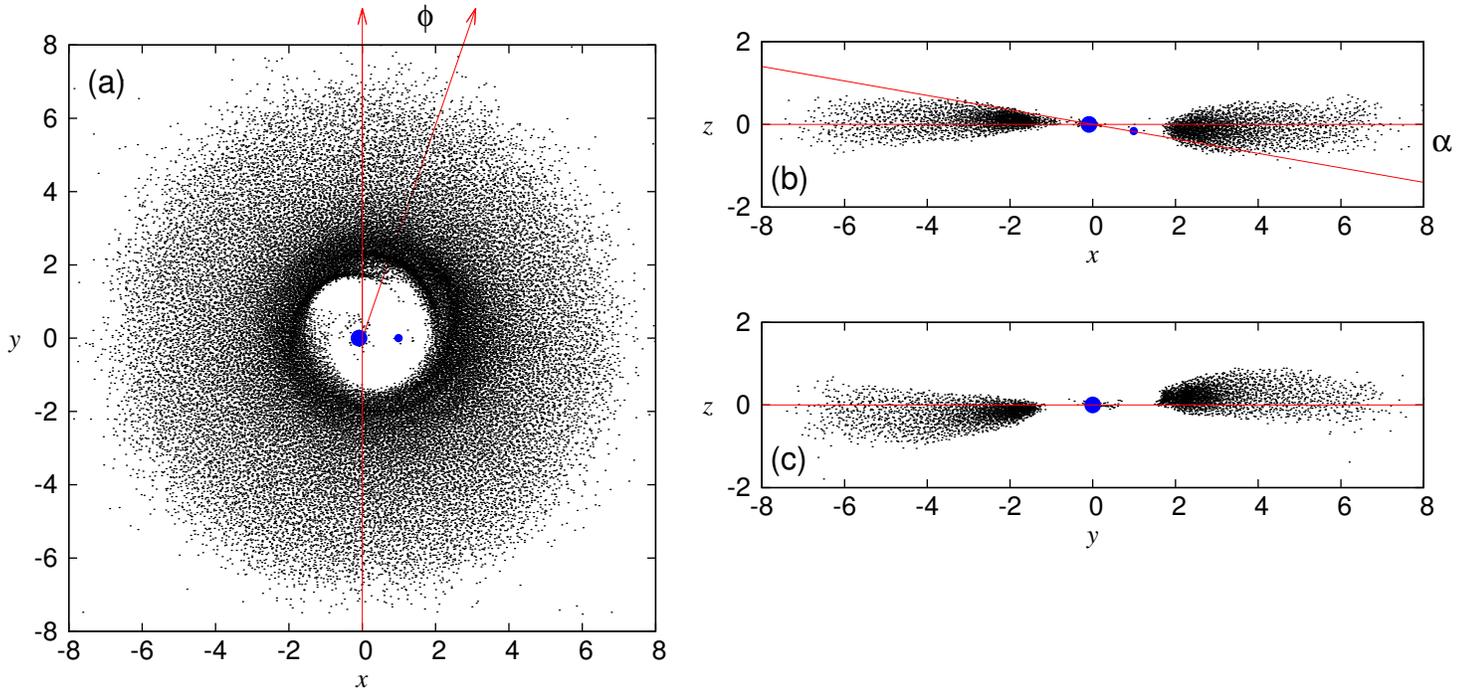}
  }
 \caption{Distribution of matter in the disk after 30 rotations of the system: (a) face-on view, (b) and (c) disk sections along the
$y$ and $x$ axes; here, $\alpha$ is the angle between the orbital plane of the binary system and the CB disk plane at the initial instant of
time. The distances are in units of the binary's semimajor axis. The model parameters are: the component mass ratio $q = 0.1$, the viscosity parameter $c = 0.05$ and $\alpha = 10^\circ$. The angle $\phi$ is measured from the $y$ axis (coincident with the line of nodes) in
the direction of the companion's rotation.}
 \label{disk}
\end{center}
\end{figure}

\clearpage
\begin{figure}[!h]\begin{center}
  \makebox[0.6\textwidth]{\includegraphics[scale=0.7]{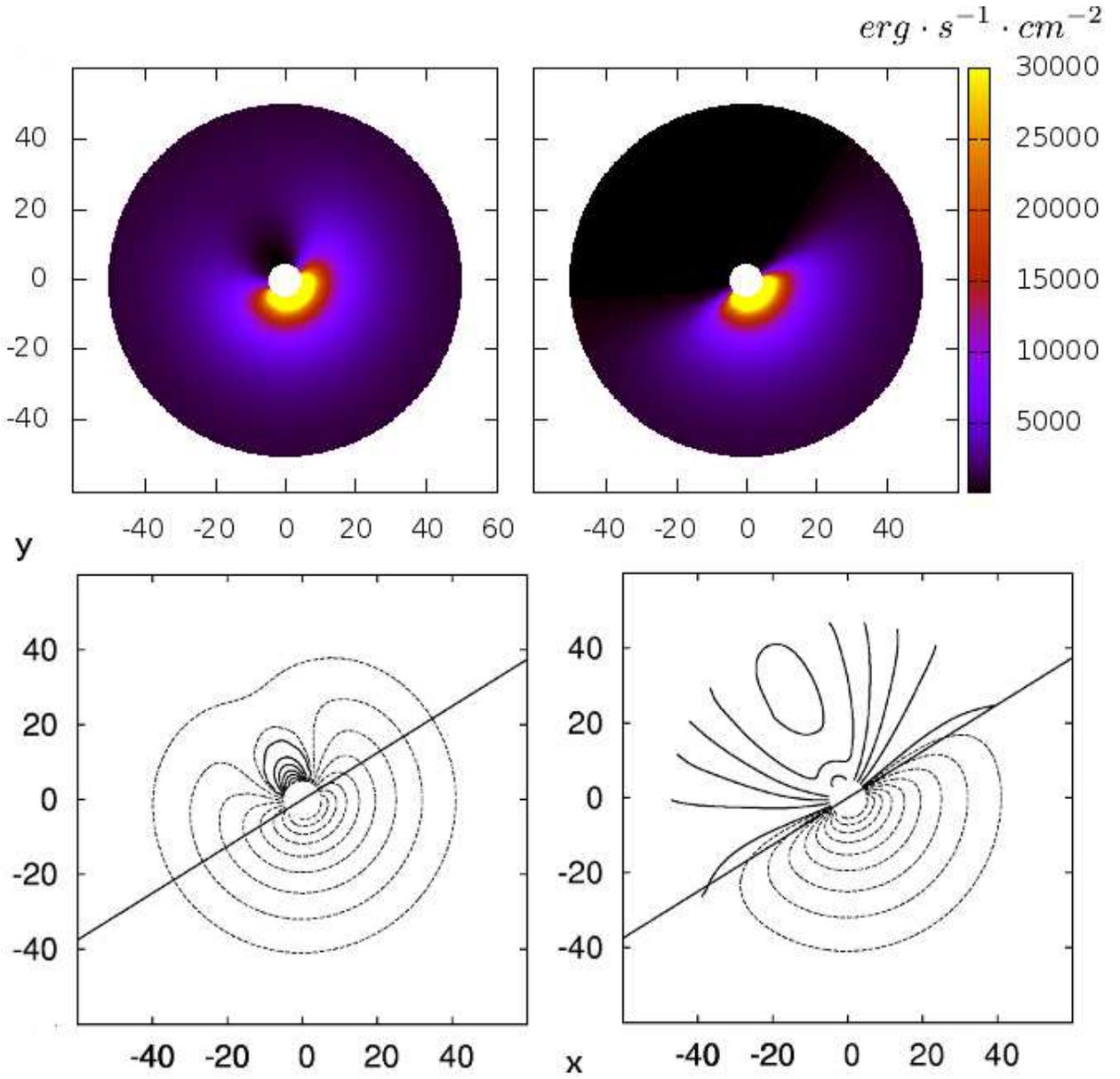}
  }
 \caption{Top: illumination of the outer part of the CB disk by the binary's primary component for two models differing by the
orbital inclination of the companion to the CB disk plane. The model parameters are: $q = 0.1$, $c = 0.05$, $\alpha = 10^\circ$ (left) and $\alpha = 30^\circ$.  (right). The readings along the $x$ and $y$ axes are in units of the binary's semimajor axis; the color scale of illumination
is in erg/(s$\times$cm$^2$). Bottom: the same in the form of isolines; the straight line corresponds to the line separating the bright and
dark (shadow) regions. The dotted lines in the lower graphs indicate the regions where the illumination is greater than that of
the outer disk at a distance of $50\,a$; the solid lines indicate the regions where it is smaller than this value.}
 \label{ill_incl}
\end{center}
\end{figure}
\begin{figure}[!h]\begin{center}
  \makebox[0.6\textwidth]{\includegraphics[scale=1.4]{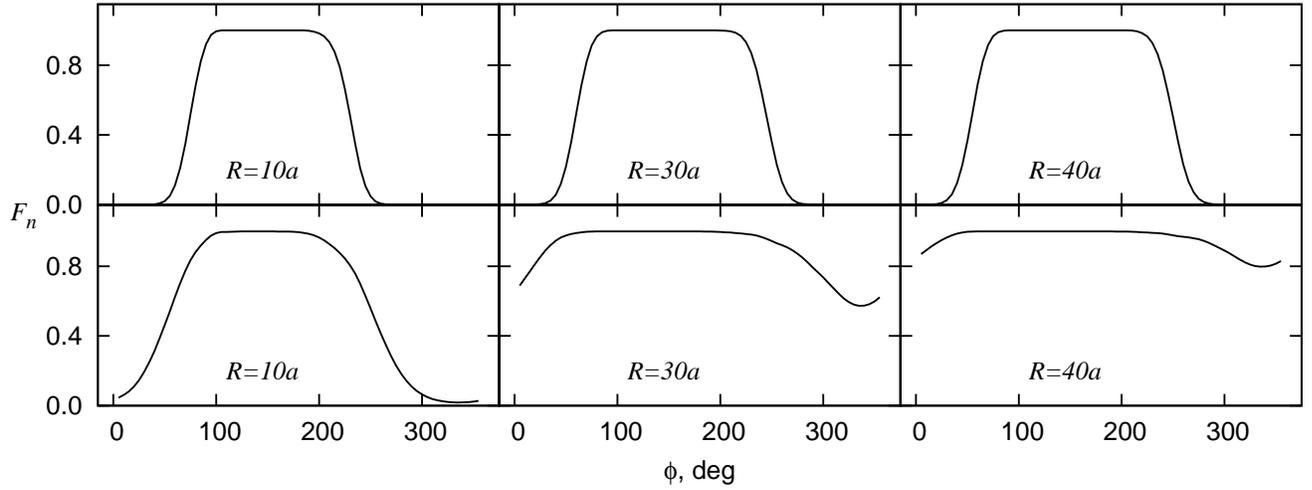}
  }
 \caption{ Normalized dependences of the disk illumination ($F_n$) on azimuth at three distances from the center (for the same
models as in Fig. 2). The distance $R$ is in units of the semimajor axis. $\alpha = 30^\circ$ (top) and $\alpha = 10^\circ$ (bottom).}
\label{F_incl}
\end{center}
\end{figure}
\begin{figure}[!h]\begin{center}
  \makebox[0.6\textwidth]{\includegraphics[scale=1.4]{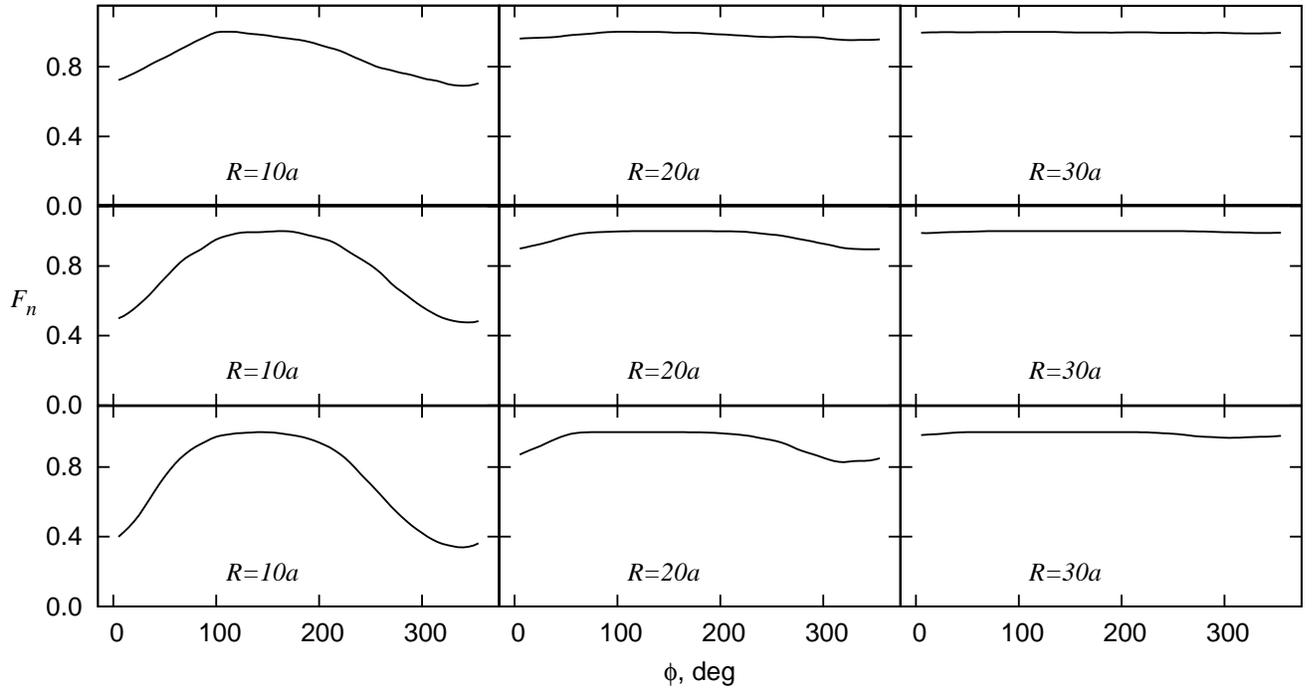}
  }
 \caption{Same as Fig. 3 for three models differing by the companion mass: $q=0.01$ (top), $q=0.03$ (middle), and $q=0.1$
(bottom). In all three models, $c=0.05$, $\alpha=5^\circ$.}
 \label{F_mass}
\end{center}
\end{figure}
\begin{figure}[!h]\begin{center}
  \makebox[0.6\textwidth]{\includegraphics[scale=1.4]{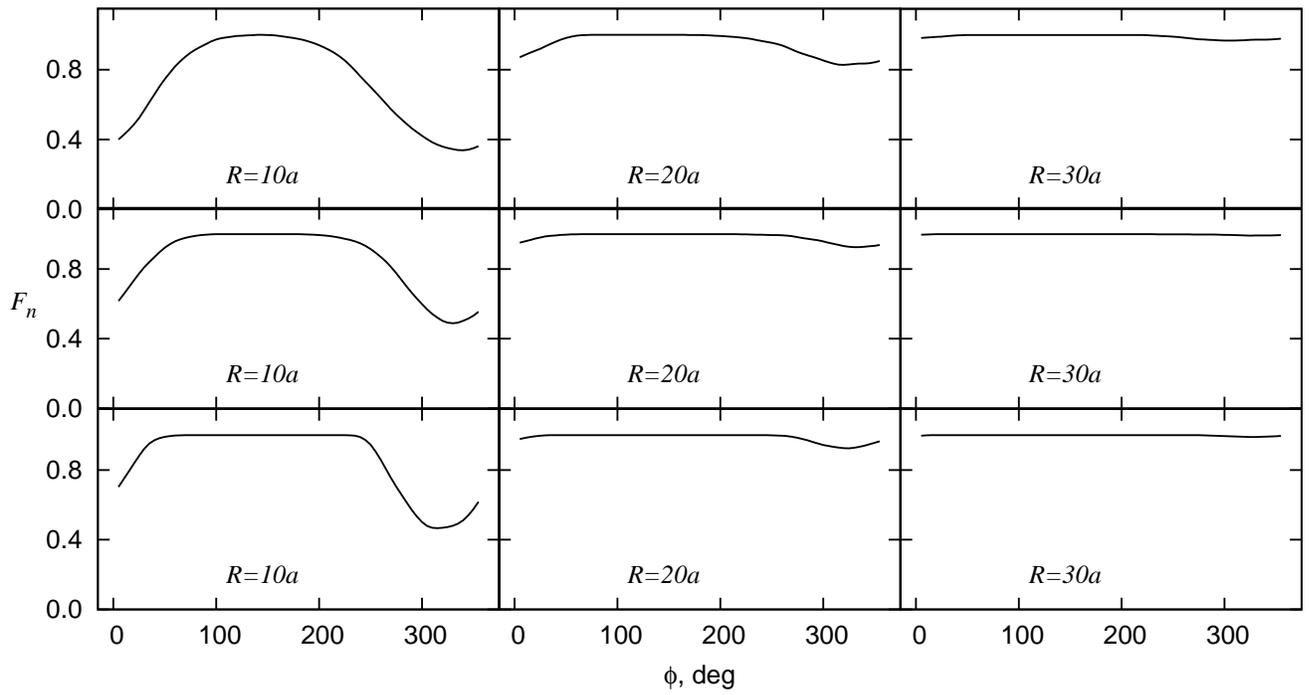}
  }
 \caption{Same as Fig. 3 for three models differing by the viscosity: the upper, middle, and lower plots correspond to $c=0.05$, $c=0.04$ and $c=0.02$, respectively. In all three models, $q = 0.1$, $\alpha = 5^\circ$.}
 \label{F_visc}
\end{center}
\end{figure}
\begin{figure}[!h]\begin{center}
  \makebox[0.6\textwidth]{\includegraphics[scale=0.7]{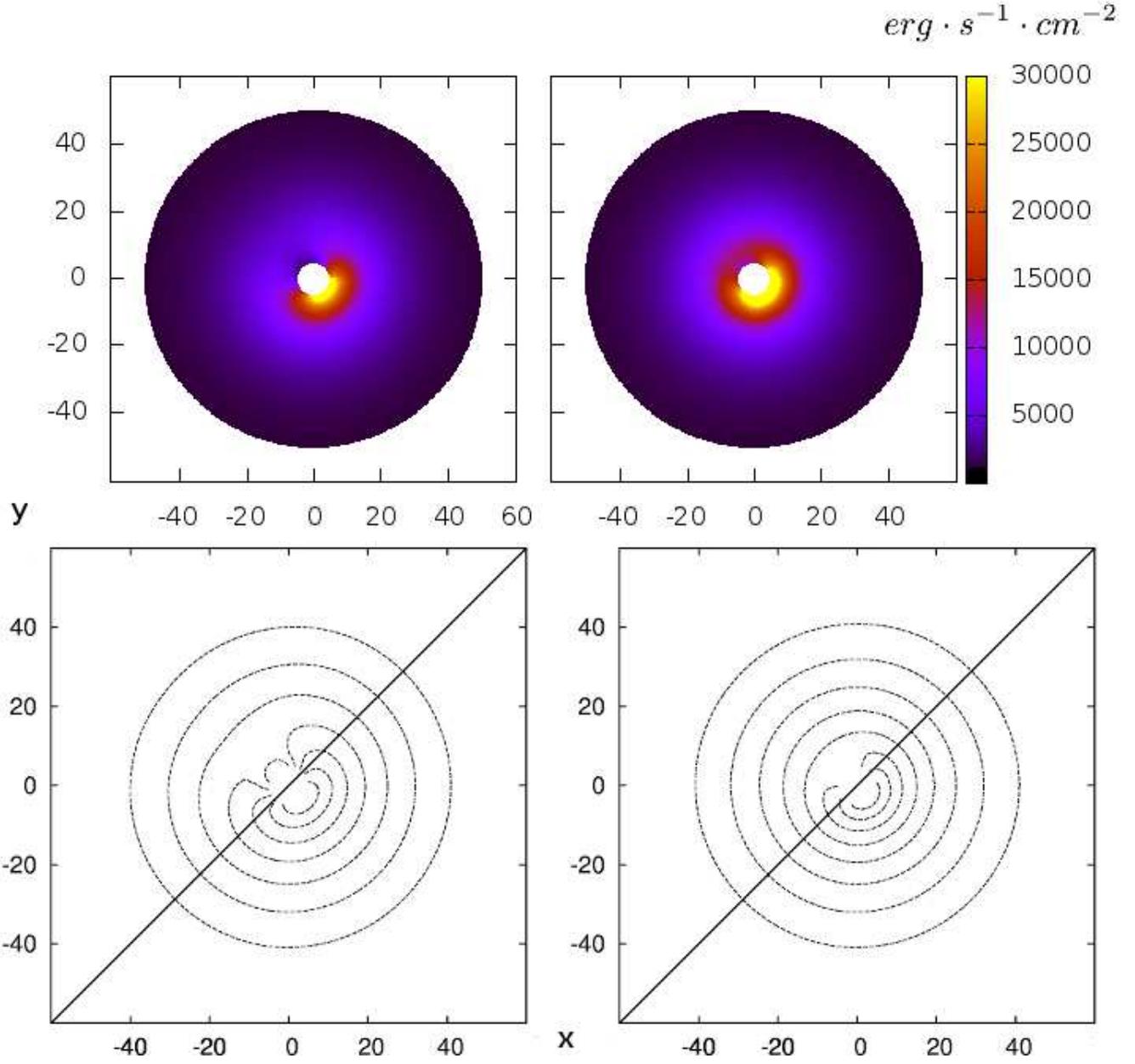}
  }
 \caption{Small perturbations in the limiting cases. Top: the disk illumination for the models with $q=0.003$, $c=0.05$, $\alpha=10^\circ$ (left) and $q=0.1$, $c=0.05$, $\alpha=3^\circ$ (right). Bottom: the same in the form of isolines; the inclination of the line separating the dark and bright regions is $\phi_l=46^\circ$ (left) and $\phi_l=44^\circ$ (right).}
\label{ill_small}
\end{center}
\end{figure}
\begin{figure}[!h]\begin{center}
  \makebox[0.6\textwidth]{\includegraphics[scale=1.4]{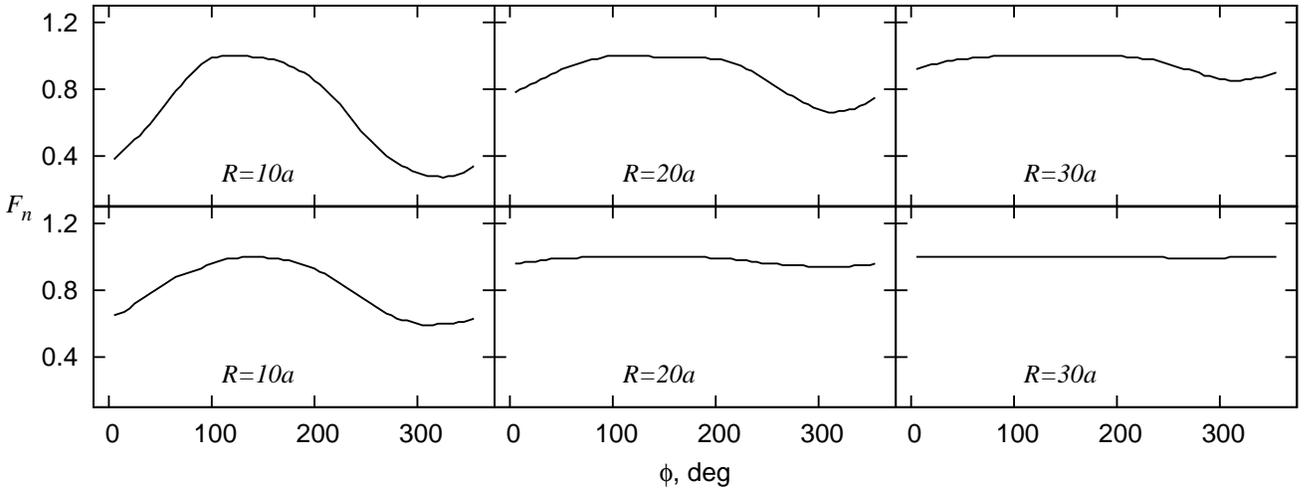}
  }
 \caption{ Same as Fig. 3 for two models with small perturbations: the upper and lower plots correspond to the models with $q = 0.003$, $c = 0.05$, $\alpha = 10^\circ$ and $q = 0.1$, $c = 0.05$, $\alpha = 3^\circ$, respectively}
 \label{F_small}
\end{center}
\end{figure}
\begin{figure}[!h]\begin{center}
  \makebox[0.6\textwidth]{\includegraphics[scale=0.7]{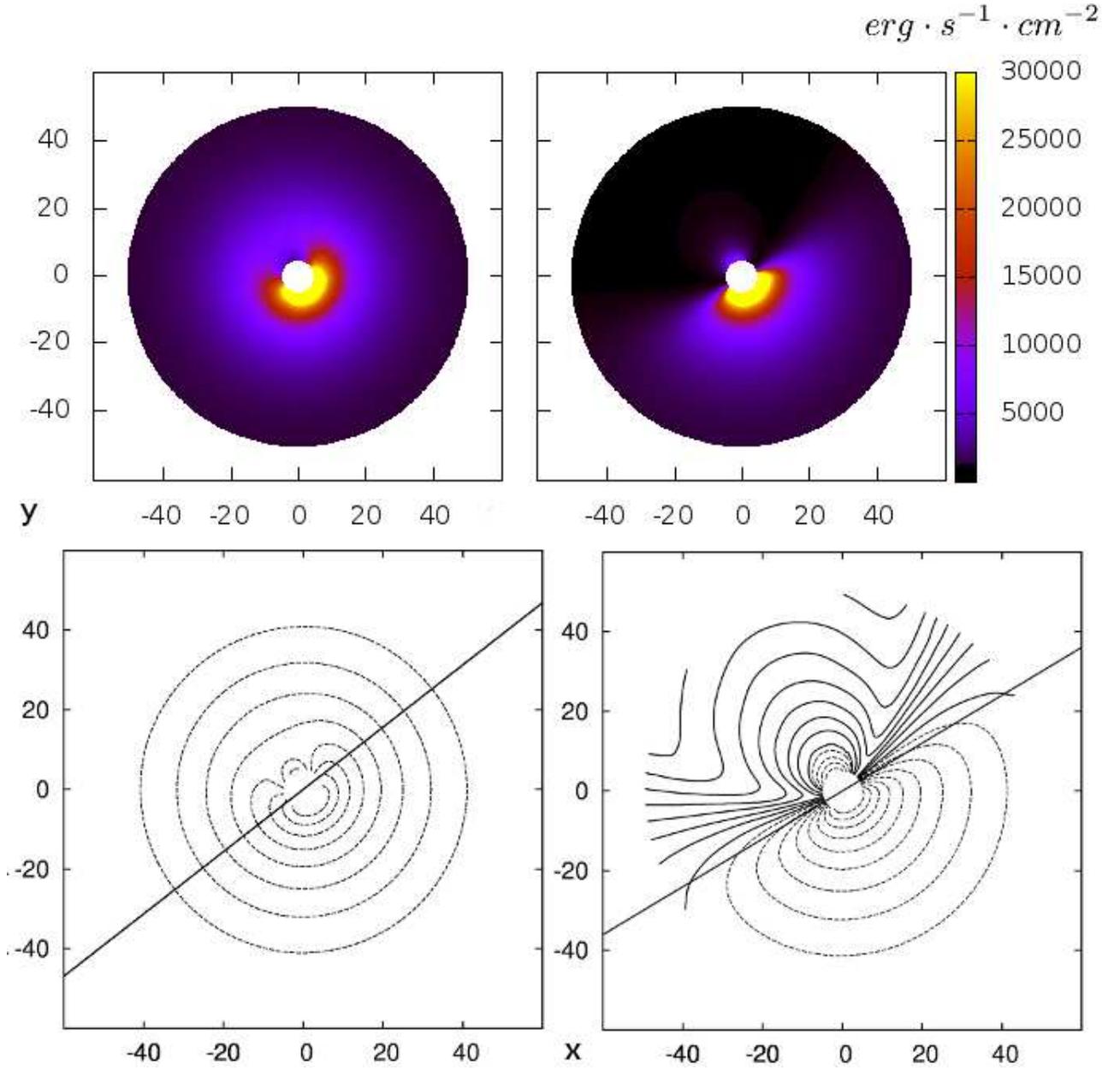}
  }
 \caption{Top: disk illumination including the scattered light for the models with $q = 0.1$, $c = 0.05$,
$\alpha =  5^\circ$ (left) and $\alpha = 30^\circ$.
 (right). Bottom: the same in the form of isolines; the inclination of the line separating the dark and bright regions is $\phi_l = 49^\circ$ (left) and $\phi_l = 59^\circ$  (right).}
 \label{ill_sca}
\end{center}
\end{figure}
\begin{figure}[!h]\begin{center}
\makebox[0.6\textwidth]{\includegraphics[scale=1.4]{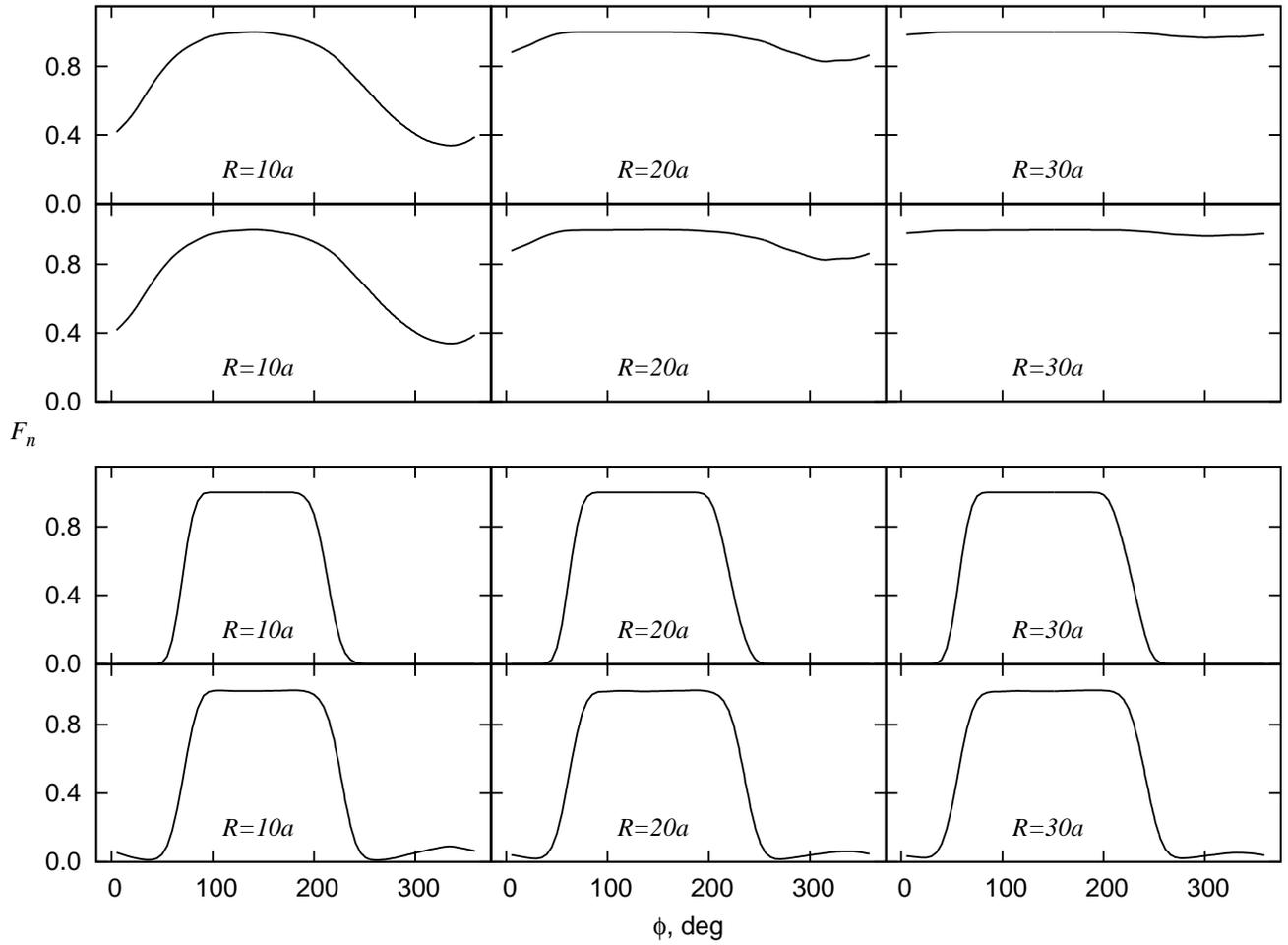} }
\caption{Same as Fig. 3 for the models with and without the scattered radiation. The model parameters are: $q=0.1$,
$c=0.05$, $\alpha =  5^\circ$ (the upper pair of plots) and $\alpha
= 30^\circ$.  (the lower pair of plots). The upper and lower plots in each pair show the
dependence $F_n$ without and with the scattered light, respectively.}
 \label{F_sca}
\end{center}
\end{figure}
\begin{figure}[!h]\begin{center}
\makebox[0.6\textwidth]{\includegraphics[scale=0.7]{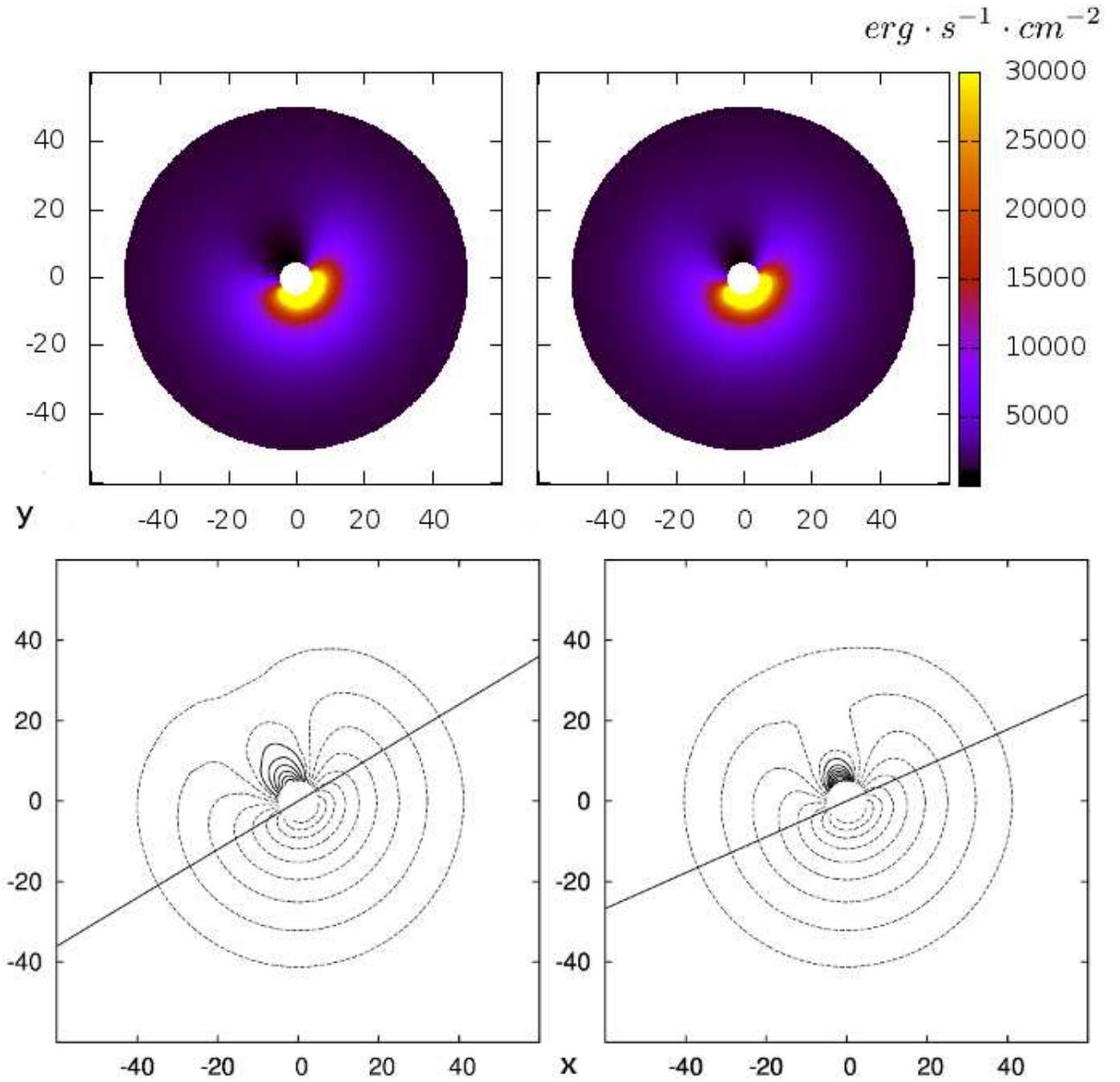}}
\caption{Illumination of the outer CB disk (with the scattered light) for two positions of the secondary components in its orbit.
Left: the companion is at $\phi = 90^\circ$ (under the disk), the inclination of the separating line is $\phi_l = 58^\circ$; right: $\phi =
270^\circ$ (the companion is above the disk), $\phi_l = 71^\circ$.
The model parameters are $q = 0.1$, $c=0.05$ end $\alpha = 10^\circ$.}
\label{orbital}
\end{center}
\end{figure}
\begin{figure}[!h]\begin{center}
\makebox[0.6\textwidth]{\includegraphics[scale=1.4]{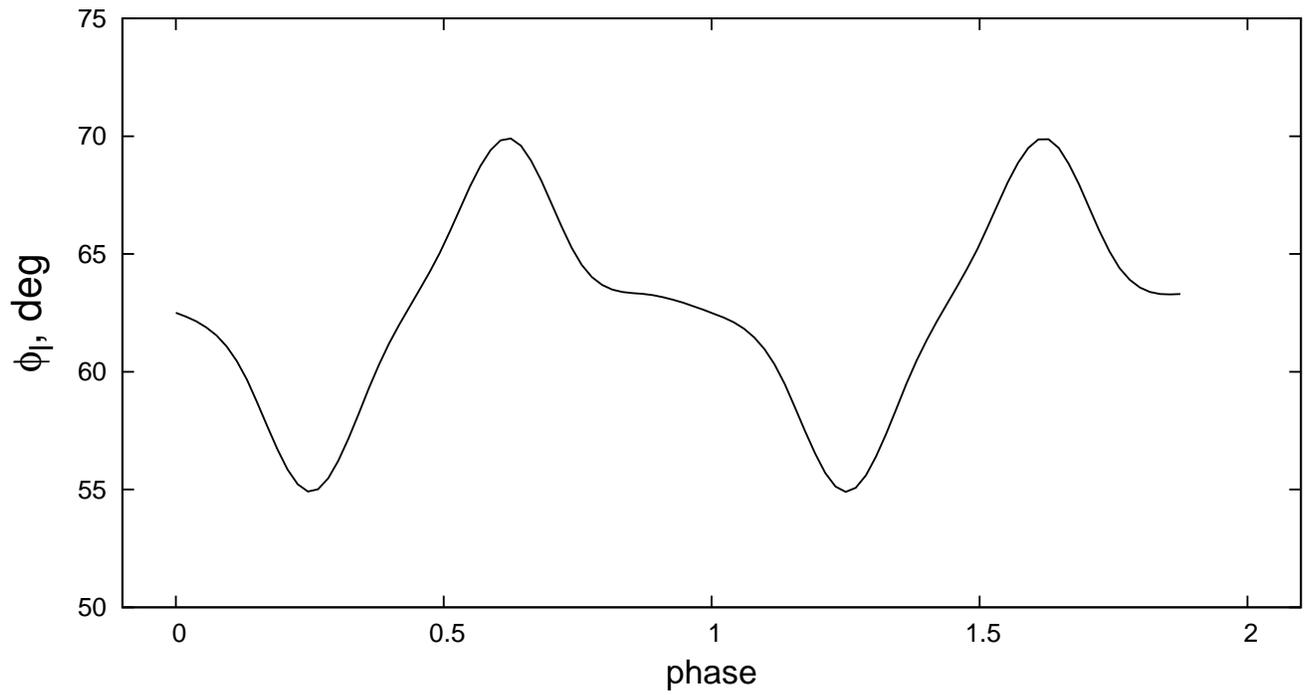}}
\caption{ Inclination ($\phi_l$) of the line separating the bright and dark regions versus orbital phase. The model parameters are
$q=0.1$, $c=0.05$ and $\alpha=10^\circ$. Phase $0$ corresponds to the position of the component at $\phi=0^\circ$.} \label{ldinclin}
\end{center}
\end{figure}
\begin{figure}[!h]\begin{center}
  \makebox[0.6\textwidth]{\includegraphics[scale=0.7]{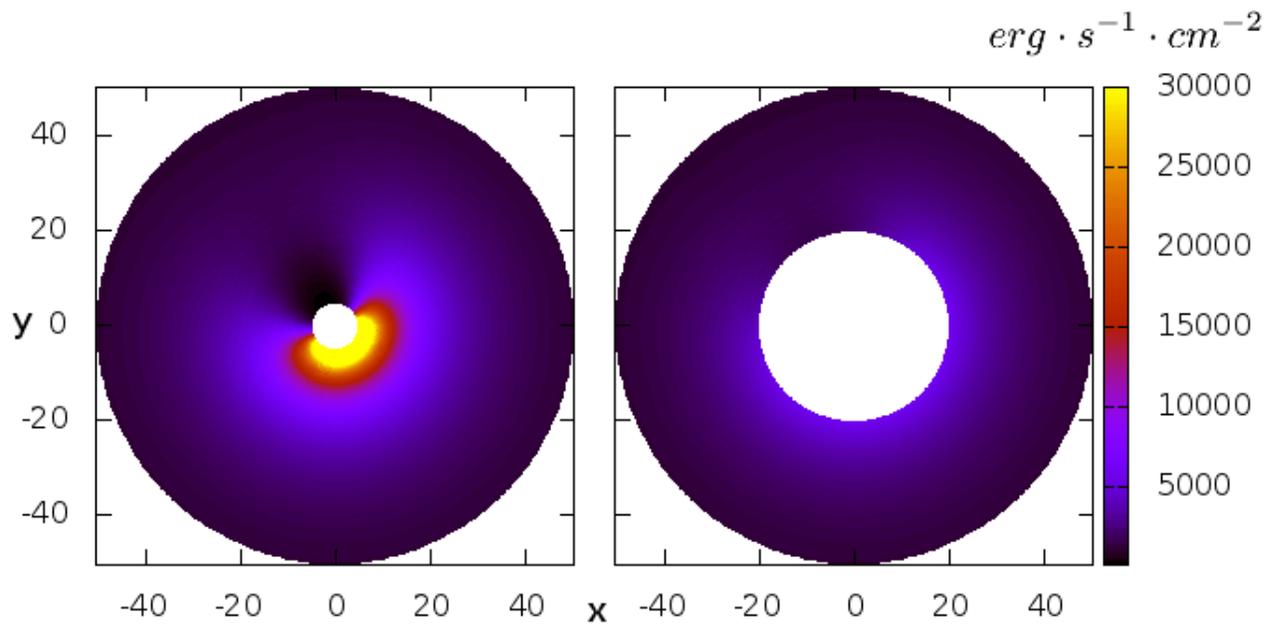}
  }
 \caption{ Illumination of the outer CB disk. The model parameters are $q=0.1$, $c=0.05$ and
 $\alpha=10^\circ$. The entire disk and only its outer region (the inner part with a size $R < 20a$
 is covered by an opaque shield) are shown on the left and the right,
respectively.}
 \label{koronograf}
\end{center}
\end{figure}

\end{document}